\documentclass[sigconf]{acmart}
\AtBeginDocument{%
  \providecommand\BibTeX{{%
    \normalfont B\kern-0.5em{\scshape i\kern-0.25em b}\kern-0.8em\TeX}}}

\copyrightyear{2022}
\acmYear{2022}
\setcopyright{acmcopyright}
\acmConference[DAC '22]{The 59th Annual Design Automation Conference 2022}{July 10--14, 2022}{San Francisco, CA, USA}
\acmBooktitle{The 59th Annual Design Automation Conference 2022 (DAC '22), July 10--14, 2022, San Francisco, CA, USA}
\acmPrice{15.00}
\acmDOI{}
\acmISBN{}

\usepackage[T1]{fontenc}
\usepackage{booktabs} 
\usepackage{subfigure}
\usepackage{multirow}
\usepackage{enumitem}
\usepackage{pifont}
\usepackage{amsmath}

\usepackage{mathtools}
\usepackage{amsfonts}
\newcommand{\overbar}[1]{\mkern 1.5mu\overline{\mkern-1.5mu#1\mkern-1.5mu}\mkern 1.5mu}
\usepackage[linesnumbered,ruled,vlined]{algorithm2e}
\usepackage{graphicx}
\usepackage{textcomp}
\usepackage{xcolor}
\usepackage{cleveref}
\crefformat{section}{\S#2#1#3}
\crefformat{subsection}{\S#2#1#3}
\crefformat{subsubsection}{\S#2#1#3}

\usepackage{nccmath}

\usepackage{titlesec}
\titlespacing\section{2pt}{2pt plus 2pt minus 1pt}{1pt plus 1pt minus 1pt}
\titlespacing\subsection{2pt}{2pt plus 2pt minus 1pt}{1pt plus 1pt minus 1pt}
\titlespacing\subsubsection{2pt}{2pt plus 2pt minus 1pt}{1pt plus 1pt minus 1pt}

\begin{document}

\title{MATCHA: A Fast and Energy-Efficient Accelerator for Fully Homomorphic Encryption over the Torus}

\author{Lei Jiang}
\authornote{This work was partially supported by NSF through awards CCF-1908992, CCF-1909509, and CCF-210597. Work done while Nrushad Joshi was at UROC@Luddy IU.}
\email{jiang60@iu.edu}
\affiliation{
\country{Indiana University}
}

\author{Qian Lou}
\email{qlou@iu.edu}
\affiliation{
\country{Indiana University}
}

\author{Nrushad Joshi}
\email{nrujoshi@iu.edu}
\affiliation{
\country{Indiana University}
}

\renewcommand{\shortauthors}{}

\begin{abstract}
Fully Homomorphic Encryption over the Torus (TFHE) allows arbitrary computations to happen directly on ciphertexts using homomorphic logic gates. However, each TFHE gate on state-of-the-art hardware platforms such as GPUs and FPGAs is extremely slow ($>0.2ms$). Moreover, even the latest FPGA-based TFHE accelerator cannot achieve high energy efficiency, since it frequently invokes expensive double-precision floating point FFT and IFFT kernels. In this paper, we propose a fast and energy-efficient accelerator, MATCHA, to process TFHE gates. MATCHA supports aggressive bootstrapping key unrolling to accelerate TFHE gates without decryption errors by approximate multiplication-less integer FFTs and IFFTs, and a pipelined datapath. Compared to prior accelerators, MATCHA improves the TFHE gate processing throughput by $2.3\times$, and the throughput per Watt by $6.3\times$. 
\end{abstract}



\maketitle

\setlength{\abovecaptionskip}{3pt plus 3pt minus 2pt}
\setlength{\textfloatsep}{3pt plus 3pt minus 2pt}

\section{Introduction}
\label{s:intro}

In cloud computing, it is dangerous for clients upload their raw data to untrusted cloud servers, due to potential data breaches. Moreover, recent legislation~\cite{Hoofnagle:ICTL2019} requires cloud computing enterprises to provide sufficient security for clients' personal data. 

Recently, \textit{Fully Homomorphic Encryption} (FHE)~\cite{Chillotti:JC2018,Brakerski:TCT2014,Cheon:CEA2020} emerges as one of the most promising cryptographic solutions to allowing arbitrary computations on encrypted data in untrusted cloud servers. Compared to Secure Multi-Party Computation, FHE requires neither frequent communications between clients and cloud servers, nor significant circuit garbling overhead on the client side. FHE enables a client to encrypt her data and to send only ciphertexts to a cloud server that can directly evaluate homomorphic functions, e.g., encrypted neural inferences~\cite{Brutzkus:ICML2019} or encrypted general-purpose computing~\cite{Matsuoka:SECURITY2021},  on the ciphertexts. When all computations are completed, the server returns the encrypted results to the client without learning any intermediate or final output, due to the end-to-end encrypted data flow. Only the client can decrypt the results by her secret key.

\textfloatsep=4pt
\begin{table}[t!]
\centering
\footnotesize
\setlength{\tabcolsep}{3pt}
\caption{The comparison between various HE schemes.}
\label{t:he_scheme_comparison}
\begin{tabular}{|c||c|c|c|}\hline
Scheme                                    & FHE Op.          & Data Type        & Bootstrapping \\\hline\hline
BGV~\cite{Brakerski:TCT2014}              & mult, add        & integer          & $\sim800s$    \\\hline
BFV~\cite{Fan:CARCH2012}                  & mult, add        & integer          & $>1000s$      \\\hline
CKKS ~\cite{Cheon:CEA2020}                & mult, add        & fixed point      & $\sim500s$    \\\hline
FHEW~\cite{DUCAS:ICTACT2015}              & Boolean          & binary           & $<1s$         \\\hline\hline
\textbf{TFHE}~\cite{Chillotti:JC2018}     & \textbf{Boolean} & \textbf{binary}  & $\mathbf{13ms}$        \\\hline
\end{tabular}
\end{table}

Among all FHE cryptosystems, FHE over the Torus (TFHE)~\cite{Chillotti:JC2018} \textit{is the most efficient scheme supporting arbitrary operations with an unlimited computation depth}, as shown in Table~\ref{t:he_scheme_comparison}. First, TFHE supports arbitrary operations by various homomorphic Boolean logic gates. Traditional FHE schemes such as BGV~\cite{Brakerski:TCT2014}, BFV~\cite{Fan:CARCH2012}, and CKKS~\cite{Cheon:CEA2020} can perform only homomorphic additions and multiplications, while both FHEW~\cite{DUCAS:ICTACT2015} and TFHE~\cite{Chillotti:JC2018} can enable homomorphic Boolean algebra, e.g., NAND, XOR, and XNOR gates. Second, TFHE obtains the fastest bootstrapping. Each FHE operation inevitably introduces a certain amount of noise into the ciphertext. If there are too many FHE operations on the computational critical path, the accumulated noise in the ciphertext may exceed a threshold, and thus the ciphertext cannot be decrypted successfully. To support an unlimited computation depth, a FHE scheme has to periodically invoke a \textit{bootstrapping} operation to decrease the amount of noise in the ciphertext. The bootstrapping operation is extremely expensive for BGV, BFV, and CKKS. For example, a BGV bootstrapping typically costs several hundred seconds~\cite{Halevi:ICTACT2015}. Therefore, these FHE schemes can support only a limited computation depth by designing a large enough noise budget. Although a bootstrapping of FHEW takes only $1s$, TFHE can obtain a even faster bootstrapping, i.e., a TFHE bootstrapping requires only $13ms$ on a CPU. By fast bootstrapping, TFHE allows an unlimited computation depth.

Unfortunately, a TFHE-based complex circuit consisting of multiple TFHE gates is still extremely slow. For instance, a TFHE-based simple RISC-V CPU~\cite{Matsuoka:SECURITY2021} comprising thousands of TFHE gates can run at only $1.25Hz$. In order to realize practical TFHE-based computing, it is critical to accelerate TFHE gates by specialized hardware. However, TFHE is only well-implemented on CPUs~\cite{Toufique:HOST2020} and GPUs~\cite{Dai:CUFHE2018}. Although a recent work~\cite{Serhan:SPSL2021} accelerates TFHE gates on a FPGA, the TFHE gate latency on the FPGA is much longer than that on a GPU. To the best of our knowledge, there is no ASIC-based hardware accelerator for TFHE.


In this paper, we propose a fast and energy-efficient accelerator, \textit{MATCHA}, to process TFHE gates. We find that the bootstrapping dominates the latency of all TFHE logic operations. The kernels of fast Fourier transform (FFT) and inverse FFT (IFFT) are the bottlenecks in a bootstrapping operation. MATCHA is designed to accelerate the TFHE bootstrapping using approximate multiplication-less integer FFTs and IFFTs. We also propose a pipelined datapath for MATCHA to support aggressive bootstrapping key unrolling~\cite{Zhou:ACCESS2018,Bourse:CRYPTO2018} that invokes FFTs and IFFTs less frequently. Our contributions can be summarized as follows.
\begin{itemize}[nosep,leftmargin=*]
\item In order to fully take advantage of the error tolerance capability of TFHE, MATCHA accelerates polynomial multiplications by approximate multiplication-less integer FFTs and IFFTs requiring only additions and binary shifts. Although approximate FFTs and IFFTs introduce errors in each ciphertext, the ciphertext can still be correctly decrypted, since the errors can be rounded off along with the noise during decryption.

\item We build a pipelined datapath consisting of TGSW clusters and external product cores to enable aggressive bootstrapping key unrolling that invokes FFTs and IFFTs less frequently during a bootstrapping operation. The datapath uses different register banks to serve sequential memory accesses during TGSW operations, and irregular memory accesses during FFTs and IFFTs.

\item We implemented, evaluated, and compared MATCHA against prior TFHE hardware accelerators. Compared to prior accelerators, MATCHA improves the TFHE gate processing throughput by $2.3\times$, and the throughput per Watt by $6.3\times$. 
\end{itemize}


\section{Background}
\label{s:intro}

\textbf{FHE}. Fully Homomorphic Encryption (FHE) enables arbitrary operations on ciphertexts. A FHE operation $\diamond$ is defined if there is another operation $\star$ such that $Dec[Enc(x_1)\diamond Enc(x_2)] = Dec[Enc(x_1\star x_2)]$, where $x_1$ and $x_2$ are input plaintexts, $Enc$ indicates encryption, and $Dec$ is decryption.

\textbf{Notation}. $\mathbb{T}$ denotes the torus of real numbers modulo 1, $\mathbb{R}/\mathbb{Z}$. For any ring $\mathcal{R}$, polynomials of the variable $X$ with coefficients in $\mathcal{R}$ are represented by $\mathcal{R}[X]$. We define $\mathbb{R}_N[X]:=\mathbb{R}[X]/(X^N+1)$, $\mathbb{Z}_N[X]:=\mathbb{Z}[X]/(X^N+1)$, and $\mathbb{T}_N[X]:=\mathbb{R}_N[X]/\mathbb{Z}_N[X]$, which are the ring of polynomials of variable $X$ with quotient $X^N + 1$ and real coefficients modulo 1. $\mathbb{B}\coloneqq\{0,1\}$ is a set, and we write vectors in bold. Given a set $\mathcal{S}$, we write $\mathbf{s}\xleftarrow{\$}\mathcal{S}$ to indicate that $\mathbf{s}$ is sampled uniformly at random from $\mathcal{S}$. We write $e \leftarrow \mathcal{X}$ to denote that $e$ is sampled according to $\mathcal{X}$.

\textbf{TFHE}. In TFHE~\cite{Chillotti:JC2018}, we assume $m\in \mathbb{B}$ is a plaintext. The encryption scheme works as follows:
\begin{itemize}[nosep,leftmargin=*]
\item $Setup(\lambda)$ first selects public parameters $n=n(\lambda)$, and $\sigma=\sigma(\lambda)$, where $\lambda$ is the security parameter. It samples and produces a secret key $\mathbf{s}\xleftarrow{\$}\mathbb{B}^n$.

\item $Enc[\mathbf{s}, m]$ samples a uniformly random vector $\mathbf{a}\xleftarrow{\$}\mathbb{T}^n$ and a noise $e\leftarrow\mathcal{D}_{\mathbb{T}_N[X],\sigma}$, where $\mathcal{D}_{\mathbb{T}_N[X],\sigma}$ is the Gaussian distribution over $\mathbb{T}_N[X]$ with a standard deviation $\sigma$. It outputs a ciphertext $(\mathbf{a},b)$, where $b=\mathbf{a}\cdot \mathbf{s}+e+m/2$.

\item $Dec[\mathbf{s}, (\mathbf{a},b)]$ returns $\lceil 2(b-\mathbf{a}\cdot \mathbf{s}) \rfloor$. It outputs plaintext correctly if the size of noise $e$ is bounded as $|e|<1/4$, since $2(b-\mathbf{a}\cdot \mathbf{s})=2e+m$, $|2e|<1/2$, and thus $\lceil 2(b-\mathbf{a}\cdot \mathbf{s}) \rfloor=m$.

\item $Logic[c_0, c_1]$ returns the ciphertext of the result of the logic operation between two ciphertexts $c_0$ and $c_1$, and the logic operation can be XOR, NAND, AND, and OR. A TFHE logic operation involves an addition between $c_0$ and $c_1$, and a bootstrapping.
\end{itemize}

\textbf{TLWE}. TLWE is a torus analogue of the learning with error (LWE) problem~\cite{Brakerski:TCT2014}. $k$ is a positive integer. $N$ is a power of 2, and $\mathcal{X}$ is a probability distribution over $\mathbb{R}_N[X]$. A TLWE secret key $\bar{\mathbf{s}}$ is a vector of $k$ polynomials over $\mathbb{Z}_N[X]$ with binary coefficients, denoted as $\bar{\mathbf{s}}\in\mathbb{R}_N[X]^k$. Given a polynomial message $\mu\in \mathbb{T}_N[X]$, a TLWE ciphertext of $\mu$ under the key $\bar{\mathbf{s}}$ is a TLWE sample $(\bar{\mathbf{a}},\bar{b})\in\mathbb{T}_N[X]^k\times \mathbb{T}_N[X]$, where $\bar{\mathbf{a}}\leftarrow \mathbb{T}_N[X]^k$ and $\bar{b}=\bar{\mathbf{s}}\cdot \bar{\mathbf{a}} + \mu +e$, where $e\leftarrow \mathcal{X}$.

\textbf{TGSW}. TGSW is the matrix extension of TLWE. Each row of a TGSW sample is a TLWE sample. An \textit{external product} $\boxdot$ that maps $\boxdot$: $TGSW \times TWLE\rightarrow TLWE$ can be defined by TFHE~\cite{Chillotti:JC2018}. The product of the TGSW ciphertext of a polynomial message $\mu_{TGSW}\in \mathbb{T}_N[X]$ and the TLWE ciphertext of a polynomial message $\mu_{TLWE}\in \mathbb{T}_N[X]$ becomes a TLWE ciphertext of a polynomial message $\mu_{TGSW}\cdot \mu_{TLWE} \in \mathbb{T}_N[X]$

\newcommand\mycommfont[1]{\footnotesize\ttfamily\textcolor{blue}{#1}}
\SetCommentSty{mycommfont}
\SetKwInput{KwInput}{Input}                
\SetKwInput{KwOutput}{Output}              

\begin{algorithm}[t!]
\DontPrintSemicolon
\KwInput{A TLWE sample $(\mathbf{a}, b)$ whose plaintext is $m_{in}$; a constant $m_{set}$; a bootstrapping key $\mathbf{BK_{s\rightarrow s''}}_{,\alpha}$; and a key-switching key $\mathbf{KS_{s'\rightarrow s}}_{,\gamma'}$ ($\mathbf{s'}=\mathbf{KeyExtract(s'')}$).}
\KwOutput{A TLWE sample encrypting $m_{out}=m_{in}\cdot m_{set}$.}
$\mu=m_{set}/2$, $\mu'=\mu/2$ \tcc{Initialization}
$\bar{b}=\lceil 2Nb\rfloor$, $\bar{a_i}=\lceil 2Na_i\rfloor$ for each $i\in[1,n]$ \label{a:line2_explain} \tcc{Rounding} 
$testv=(1+X+\ldots+X^{N+1})\cdot X^{N/2}\cdot \mu'$\\
$ACC\leftarrow X^{\bar{b}}\cdot (0, testv)$ \tcc{$ACC=TLWE(X^{(\bar{b}-\bar{a}s)}\cdot testv)$}
\For{$i=1$ to $n$} 
{
  $\mathbf{BK}_i = \mathbf{h} + (X^{-\bar{a_i}}-1)\cdot \mathbf{BK}_i$\\
  $ACC\leftarrow \mathbf{BK}_i \boxdot ACC$ \label{a:line6_accum} \tcc{BlindRotate}
}
$\mathbf{u}=(0,\mu') + SampleExtract(ACC)$ \tcc{Extract}
\Return $KeySwitch_{KS}(\mathbf{u})$ \tcc{KeySwitch}
\caption{The bootstrapping operation of TFHE.}
\label{a:feta_bootstrapping_alg}
\end{algorithm}

\textbf{Bootstrapping}. Each TFHE logic operation inevitably introduces a certain amount of noise into the resulting ciphertext. A bootstrapping has to be performed to remove the noise at the end of each TFHE logic operation. In various TFHE logic operations, the bootstrapping step is the largest performance bottleneck. The details of a TFHE bootstrapping can be viewed in~\cite{Chillotti:JC2018}. The bootstrapping procedure is shown in Algorithm~\ref{a:feta_bootstrapping_alg}. The dimension of the TLWE sample is set as $k=1$~\cite{Chillotti:JC2018}, which means that the TLWE sample is simply the Ring-LWE sample $(\bar{a},\bar{b})\in \mathbb{T}_N[X]\times \mathbb{T}_N[X]$. The most computationally intensive step of a bootstrapping is the homomorphic decryption in line~\ref{a:line6_accum}, where the message of $ACC$ becomes a polynomial $X^{\bar{b}-\bar{\mathbf{a}}\mathbf{s}}\cdot testv$. Particularly, homomorphically computing $X^{-\bar{\mathbf{a}}\mathbf{s}}=X^{\sum_{i=1}^n-\bar{\mathbf{a}_i}\mathbf{s}_i}=\prod_{i=1}^n X^{-\bar{\mathbf{a}_i}\mathbf{s}_i}$ involves a great number of polynomial multiplications. Na\"ively multiplying two degree $N$ polynomials has the complexity of $\mathcal{O}(N^2)$. FFT and IFFT are used to reduce the complexity of a polynomial multiplication to $\mathcal{O}(N \log(N))$~\cite{Dai:CUFHE2018}, where $N$ is the degree of polynomials.

\begin{figure*}[t!]
\begin{minipage}{.3\textwidth}
\includegraphics[width=2in]{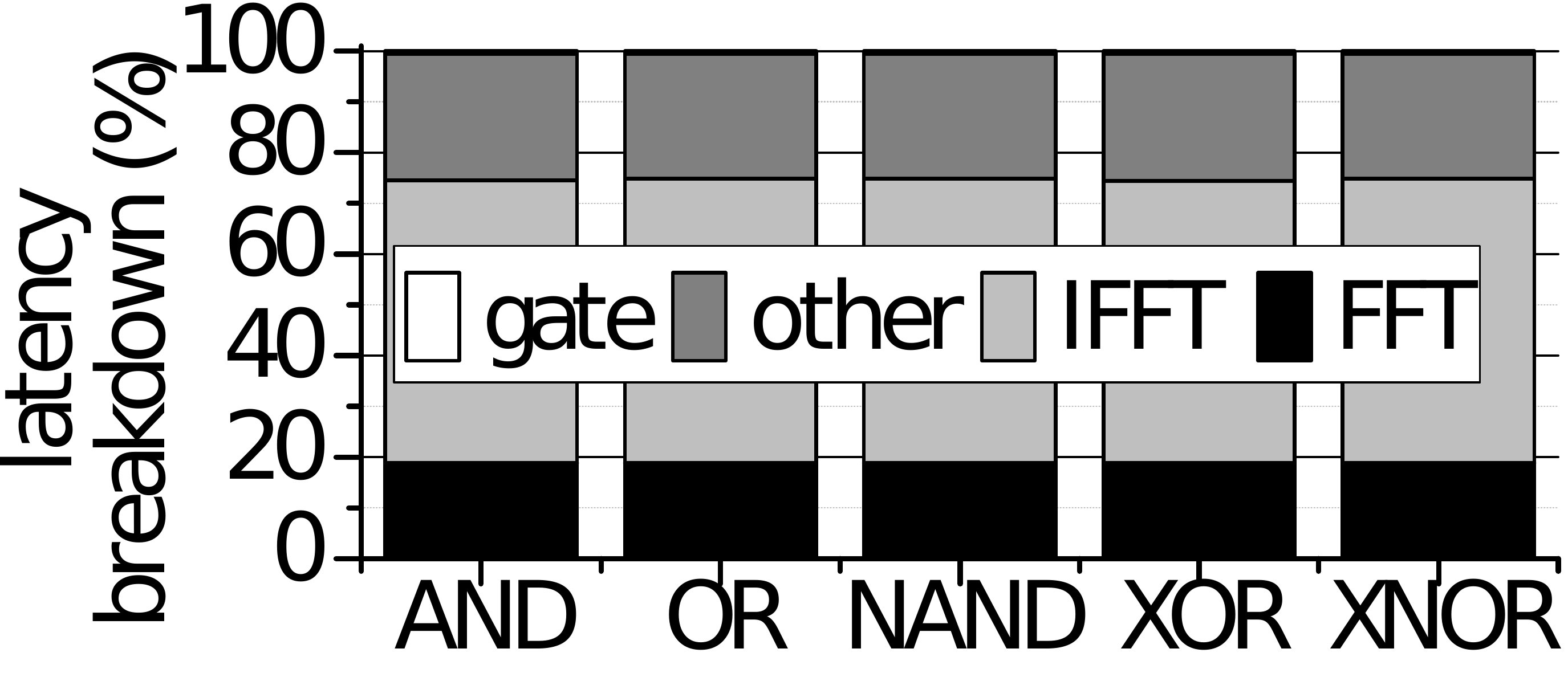}
\caption{Latency breakdown.}
\label{f:mat_kernel_break}
\end{minipage}
\hspace{-0.1in}
\begin{minipage}{.25\textwidth}
\centering
\includegraphics[width=1.6in]{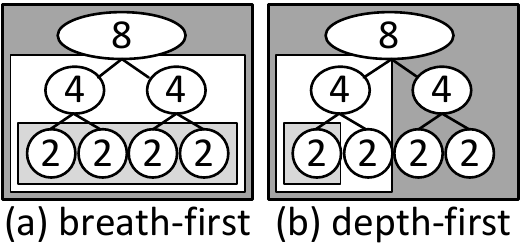}
\vspace{0.05in}
\caption{The depth-first FFT.}
\label{f:mat_depth_first}
\end{minipage}
\hspace{-0.1in}
\begin{minipage}{.45\textwidth}
\centering
\includegraphics[width=3in]{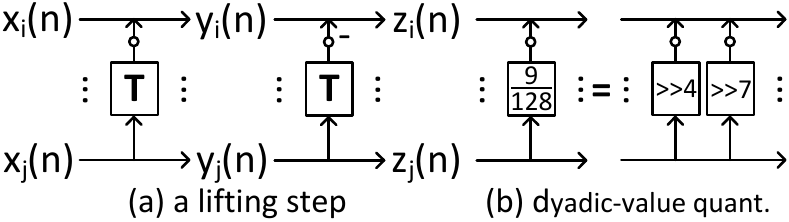}
\caption{The lifting butterfly w/o multiplication.}
\label{f:mat_lifting_op}
\end{minipage}
\vspace{-0.2in}
\end{figure*}

\textbf{Torus Implementation}. Theoretically, the scale invariant scheme of TFHE is defined over the real torus $\mathbb{T}$, where all operations are modulo 1. But TFHE rescales the elements over $\mathbb{T}$ by a factor $2^{32}$, and maps them to 32-bit integers~\cite{Chillotti:JC2018}, since it can work with approximations. Therefore, TFHE does not have to actively perform modular reduction, since all operations on 32-bit integers implicitly call a native and automatic mod $2^{32}$ operation. To maintain high conversion accuracy, TFHE uses 64-bit double-precision floating point FFT and IFFT kernels~\cite{Chillotti:JC2018}.

\section{Related Work and Motivation}
\label{s:related}

\textbf{Related Work}. Except some TFHE implementations on CPUs~\cite{Chillotti:JC2018}, GPUs~\cite{Dai:CUFHE2018}, and FPGAs~\cite{Serhan:SPSL2021}, there is no specialized hardware accelerator that can process TFHE. A TFHE accelerator is different from the accelerators designed for other FHE schemes such as BGV, BFV, and CKKS in two points. First, although few prior accelerators~\cite{Feldmann:MICRO2021} support BGV and CKKS bootstrapping along a tiny multiplicative depth datapath, most prior works~\cite{Riazi:ASPLOS2020,Roy:HPCA2019,Ahmet:DATE2020} design hardware accelerators to process leveled BFV or CKKS homomorphic operations without bootstrapping. However, a TFHE accelerator must perform bootstrapping at the end of each TFHE gate. Second, BGV, BFV, and CKKS require NTT and INTT kernels, while TFHE needs only FFT and IFFT kernels without modular reduction.

\textbf{Motivation}. A TFHE gate performs not only polynomial additions but also a bootstrapping (FFT+IFFT+other) that costs 99\% of the gate latency on a CPU, as shown in Figure~\ref{f:mat_kernel_break}. Therefore, in order to shorten the latency of TFHE gates, we need to accelerate the bootstrapping step in TFHE gates. Moreover, FFTs and IFFTs consume 80\% of the bootstrapping latency in various TFHE gates. In order to accelerate TFHE gates, MATCHA adopts approximate multiplication-less integer FFTs and IFFTs, and uses a pipelined datapath to support aggressive bootstrapping key unrolling~\cite{Zhou:ACCESS2018,Bourse:CRYPTO2018}.


\section{MATCHA}
\label{s:matcha}

\subsection{Approximate Fast Integer FFT and IFFT}

Despite the fact that elements over $\mathbb{T}$ are mapped to 32-bit integers, TFHE still uses 64-bit double-precision floating point FFT and IFFT kernels, since 32-bit integer or single-precision floating point FFT and IFFT kernels are not accurate enough to guarantee the correct decryption of a ciphertext~\cite{Chillotti:JC2018}. However, processing 64-bit double-precision floating point FFT and IFFT kernels incurs significant hardware overhead and power consumption.

\textbf{Novelty}. We first identify the opportunity to use approximate integer FFTs and IFFTs to accelerate TFHE without decryption errors for MATCHA. It is difficult to apply approximate NTTs and INTTs in accelerating other FHE schemes, e.g., BGV, BFV, and CKKS, which do not include a bootstrapping step after each homomorphic multiplication or addition. The errors introduced by approximate NTTs and INTTs will be quickly accumulated in the ciphertext and result in a decryption error, if a bootstrapping step cannot be performed in time. On the contrary, TFHE keeps the approximation errors of integer FFTs and IFFTs in check by performing a bootstrapping step at the end of each TFHE gate.



\textbf{Depth-first FFT}. Most prior FHE accelerators~\cite{Feldmann:MICRO2021,Riazi:ASPLOS2020,Roy:HPCA2019} perform NTTs and INTTs by the Cooley-Tukey data flow that introduces irregular memory accesses particularly in its bit-reversal stage. In order to remove the bit-reversal overhead, a prior ideal-lattice-based cryptographic accelerator~\cite{Liu:TECS2017} uses the Cooley-Tukey flow for NTTs and the Gentlemen-Sande flow for INTTs. These cryptographic accelerators store a polynomial mod $X^N+1$ as a list of $N$ coefficients. For each multiplication between two polynomials, they execute two NTT kernels on two polynomials respectively, perform element-wise multiplications, and then run an INTT kernel on the result. The invoking frequency ratio between NTTs and INTTs is $2:1$. These FHE accelerators have are many opportunities (i.e., switchings from NTT to INTT) to reduce the bit-reversal overhead. In contrast, TFHE saves a polynomial mod $X^N+1$ as either a list of $N$ coefficients or the Lagrange half-complex representation consisting in the complex evaluations of the polynomial over the roots of unity $exp(i(2j+1)\pi/N)$ for $j\in\llbracket0,\frac{N}{2}\llbracket$. FFT and IFFT kernels are required only during the conversion between these two representations. The invoking frequency ratio between FFTs and IFFTs in a TFHE gate is $1:4$. As Figure~\ref{f:mat_kernel_break} shows, the latency of IFFT kernels is much longer than FFT kernels. TFHE does not have many opportunities to reduce the bit-reversal overhead. Instead, for MATCHA, we focus on decreasing the computing overhead of a single FFT or IFFT kernel. We adopt the depth-first iterative conjugate-pair FFT (CPFFT) algorithm~\cite{Becoulet:TSP2021}. Unlike the Cooley-Tukey or Gentlemen-Sande flow, the CPFFT requires only a single complex root of unity read per radix-4 butterfly. Two butterflies in the same block can share the same twiddle factor, further halving the number of reads to the twiddle-factor buffer~\cite{Becoulet:TSP2021}. Moreover, the Cooley-Tukey and Gentlemen-Sande flows process FFTs/IFFTs stage by stage in a breadth-first manner, as shown in Figure~\ref{f:mat_depth_first}(a). To capture the spatial locality, as Figure~\ref{f:mat_depth_first}(b) shows, CPFFT traverses the FFT flow in a depth-first fashion by completing a sub-transform before moving to the next.



\textbf{A Multiplication-less Butterfly}. The lifting structure~\cite{Oraintara:TSP2002}, a special type of lattice substrate implemented by cascading identity matrices with a single nonzero off-diagonal element, is proposed to approximate multiplications in FFT and IFFT kernels by additions and binary shifts. The basic lifting step shown in Figure~\ref{f:mat_lifting_op}(a) can be expressed by $y_j(n) = x_j(n)$, $y_i(n) = x_i(n) + \lceil T_{x_j}(n)\rfloor$, $z_j(n) = y_j(n)$, and $z_i(n) = y_i(n) - \lceil T_{y_j}(n)\rfloor$, where $T$ is a lifting coefficient. And thus, the lifting structure with the rounding operation can achieve integer-to-integer transform. Also, the lifting and its inverse matrices in this case are represented as
$\begin{bmatrix}
1 & T\\
0 & 1
\end{bmatrix}$ and 
$\begin{bmatrix}
1 & T\\
0 & 1
\end{bmatrix}^{-1}=
\begin{bmatrix}
1 & -T\\
0 & 1
\end{bmatrix}$, respectively. A floating-point lifting coefficient can be quantized as an approximate dyadic-valued coefficient $\alpha/2^\beta$, and hence computed with only adders and shifters, where we allocate $\beta$ bits to the lifting coefficient, and $\alpha, \beta \in \mathbb{N}$. For example, a coefficient $9/128$ can be operated as $\frac{9}{128}=\frac{2^3+2^0}{2^7}=\frac{1}{2^4}+\frac{1}{2^7}$. Hence, the lifting with its coefficient $9/128$ and a rounding operation is replaced to the summation of 4 and 7 bit-shifters illustrated in Figure~\ref{f:mat_lifting_op}(b). The perfect reconstruction in lifting is always kept if floating-point coefficients are approximated to dyadic-valued coefficients. 


\begin{figure}
\centering
\begin{minipage}{0.4\linewidth}
\centering
\includegraphics[width=1.1in]{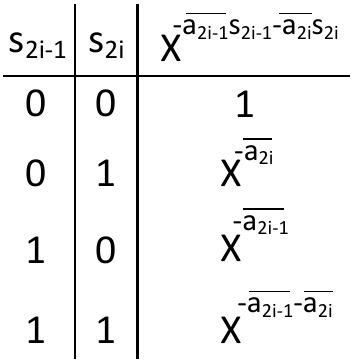}
\caption{The truth table of $X^{-\overbar{a_{2i-1}}\cdot s_{2i-1}-\overbar{a_{2i}}\cdot s_{2i}}$.}
\label{f:mat_boot_table}
\end{minipage}
\hspace{0.2in}
\begin{minipage}{0.4\linewidth}
\centering
\includegraphics[width=1.3in]{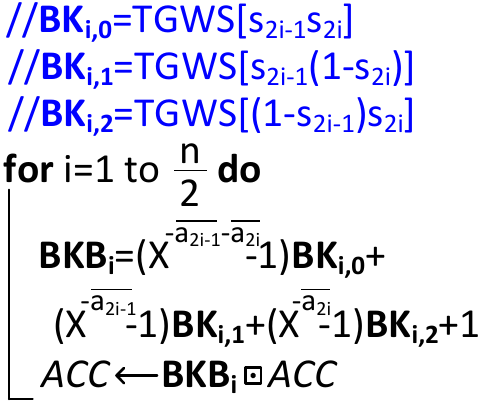}
\caption{Bootstrapping key unrolling.}
\label{f:mat_boot_alg}
\end{minipage}
\vspace{-0.15in}
\end{figure}

\begin{figure}[t!]
\centering
\includegraphics[width=3.3in]{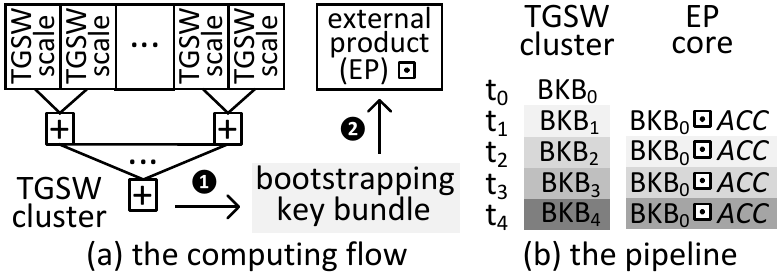}
\caption{The pipelined MATCHA for aggressive BKU.}
\label{f:mat_pipeline_hardware2}
\end{figure}

\vspace{-0.05in}
\subsection{Aggressive Bootstrapping Key Unrolling}

\textbf{Bootstrapping Key Unrolling}. A TFHE bootstrapping needs to compute external produces, i.e., $X^{-\overbar{\mathbf{a}}\mathbf{s}}=X^{\sum_{i=1}^n-\overbar{\mathbf{a}_i}\mathbf{s}_i}$ sequentially, thereby becoming the performance bottleneck of a TFHE gate. Instead, bootstrapping key unrolling (BKU)~\cite{Zhou:ACCESS2018,Bourse:CRYPTO2018} is proposed to compute $X^{\sum_{i=1}^{n/2}-\overbar{\mathbf{a}_{2i-1}}\mathbf{s}_{2i-1}-\overbar{\mathbf{a}_{2i}}\mathbf{s}_{2i}}$ in each external product, so that the number of homomorphic additions can be reduced from $n$ to $n/2$. The secret key $\mathbf{s}$ is sampled from $\mathbb{B}^n$, so $s_i\in\{0,1\}$, where $0\leq i \leq n$. Based on the values of $s_{2i}$ and $s_{2i+1}$, the truth table of $X^{\sum_{i=1}^{n/2}-\overbar{\mathbf{a}_{2i-1}}\mathbf{s}_{2i-1}-\overbar{\mathbf{a}_{2i}}\mathbf{s}_{2i}}$ can be shown in Figure~\ref{f:mat_boot_table}. So BKU rewrites $X^{-\overbar{a_{2i-1}}\cdot s_{2i-1}-\overbar{a_{2i}}\cdot s_{2i}}$ as $X^{-\overbar{a_{2i-1}}-\overbar{a_{2i}}} \cdot s_{2i-1} s_{2i} - X^{-\overbar{a_{2i-1}}} \cdot s_{2i-1}(1-s_{2i}) - X^{-\overbar{a_{2i}}} \cdot (1-s_{2i-1})s_{2i} - (1-s_{2i-1})(1-s_{2i})$. Due to the fact that $s_{2i-1}s_{2i}+ (1-s_{2i})s_{2i-1} + s_{2i}(1-s_{2i-1}) + (1-s_{2i-1})(1-s_{2i})$ is always equal to 1~\cite{Bourse:CRYPTO2018}, $X^{-\overbar{a_{2i-1}}\cdot s_{2i-1}-\overbar{a_{2i}}\cdot s_{2i}}$ can be further simplified to $(X^{-\overbar{a_{2i-1}}-\overbar{a_{2i}}}-1) \cdot s_{2i-1} s_{2i} + (X^{-\overbar{a_{2i-1}}}-1) \cdot s_{2i-1}(1-s_{2i}) - (X^{-\overbar{a_{2i}}}-1) \cdot (1-s_{2i-1})s_{2i} +1$. As Figure~\ref{f:mat_boot_alg} shows, BKU encrypts $s_{2i-1} s_{2i}$, $s_{2i-1}(1-s_{2i})$, and $(1-s_{2i-1})s_{2i}$ as TGSW ciphertexts, and builds a bootstrapping key bundle to unroll the orginal bootstrapping key for two times.


\textbf{Aggressive BKU Performing Badly on CPUs}. BKU can be further generalized as 
\begin{equation}
X^{\sum_{i=1}^{\frac{n}{m}}-\overbar{\mathbf{a}_{m\cdot i}}\mathbf{s}_{m\cdot i}-\overbar{\mathbf{a}_{m\cdot i+1}}\mathbf{s}_{m\cdot i+1}-\ldots-\overbar{\mathbf{a}_{m\cdot i+m-1}}\mathbf{s}_{m\cdot i+m-1}},
\label{e:mat_bku_aggre}
\end{equation}
where $m\in[2,n]$. So it is possible to more aggressively unroll the bootstrapping key by increasing $m$. Although unrolling the bootstrapping key for two times ($m=2$) reduces the bootstrapping latency by 49\%, we find that further enlarging $m$ beyond 2 even prolongs the bootstrapping latency on a CPU, as explained in Section~\ref{s:result}. Our experimental methodology is described in Section~\ref{s:method}. The reason can be summarized as follows.
\begin{itemize}[nosep,leftmargin=*]
\item \textbf{The limited number of cores on a CPU}. With an enlarged $m$, there are more terms in the exponent part of Equation~\ref{e:mat_bku_aggre}. For instance, when $m=4$, there are 15 terms, each of which requires a TGSW scale-and-add operation. Unfortunately, our CPU baseline has only 8 physical cores. Mapping each terms to a core, and summing the results from all cores introduce significant communication overhead.

\item \textbf{More cache conflicts}. The size of bootstrapping key increases exponentially with an enlarged $m$. For example, as Figure~\ref{f:mat_boot_alg} shows, instead of a single bootstrapping key, BKU with $m=2$ requires three bootstrapping keys. Each TGSW scale-and-add operation happening on a term fetches its corresponding bootstrapping key to the shared last level cache, generating more cache conflicts.

\item \textbf{The lack of a pipelined design}. As Figure~\ref{f:mat_boot_alg} highlights, in each iteration, the construction of the bootstrapping key bundle $\mathbf{BKB}$ and the external product operation are executed sequentially. Although it is possible to start the computation of $\mathbf{BKB}$ for the next iteration and perform the external product operation of this iteration at the same time, the current BKU implementation~\cite{Zhou:ACCESS2018} cannot do this, due to the lack of a pipelined design.  
\end{itemize}

\begin{figure}[t!]
\centering
\includegraphics[width=3.3in]{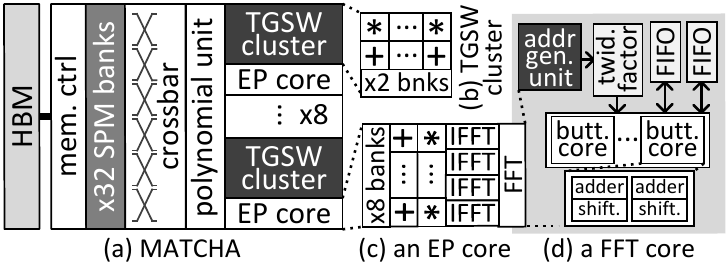}
\caption{The architecture of MATCHA (mem. ctrl: memory controller; addr gen.: address generation; twid: twiddle factor; butt.: butterfly; and shift.: shifter).}
\label{f:mat_overall_arch}
\end{figure}

\textbf{MATCHA for Aggressive BKU}. In this paper, we propose a pipeline flow for MATCHA to support aggressive BKU with a larger $m$. Compared to our CPU baseline, our pipeline flow can be easily accelerated by a large number of specialized hardware components including TGSW clusters and External Product (EP) cores. As Figure~\ref{f:mat_pipeline_hardware2}(a) shows, we divide the bottleneck of a TFHE bootstrapping into two steps, i.e., the construction of the bootstrapping key bundle, and the EP operation. A TGSW cluster is used to construct the bootstrapping key bundle, while an EP core processes EP operations between the bootstrapping key bundle and $ACC$. \ding{182} A TGSW cluster consists of a TGSW adder tree and multiple TGSW scale units, each of which computes one term in the bootstrapping key bundle, e.g., when $m=2$, $(X^{-\overbar{a_{2i-1}}-\overbar{a_{2i}}}-1)\cdot \mathbf{BK_{i,0}}$, where $\mathbf{BK_{i,0}}$ is the TGSW ciphertext of $s_{2i-1} s_{2i}$. And then, the TGSW adder sums all terms and generates the bootstrapping key bundle. \ding{183} With the bootstrapping key bundle ($\mathbf{BKB_i}$), an EP core computes $ACC\leftarrow\mathbf{BKB_i}\boxdot ACC$. The TGSW cluster and the EP core have their separated register file banks to reduce on-chip memory conflicts. Moreover, these two steps of a TFHE bootstrapping can be deployed on a TGSW cluster and an EP core in a pipelined manner, as shown in Figure~\ref{f:mat_pipeline_hardware2}(b). In each time step, the EP core computes the EP operation with the bootstrapping key bundle generated by the TGSW cluster in the previous time step. When $m$ is increased, the workload of the bootstrapping key bundle construction becomes larger. The workloads of the two steps in the pipeline can be approximately balanced by adjusting $m$.

\subsection{The Architecture of MATCHA}

\textbf{Architecture}. The overall architecture of MATCHA is shown in Figure~\ref{f:mat_overall_arch}(a). MATCHA has multiple computing components including a polynomial unit, eight TGSW clusters, and eight External Product (EP) cores. All computing components of MATCHA are connected to 32 scratchpad memory (SPM) banks by crossbars. MATCHA also employs a memory controller to manage the off-chip memory requests issued to HBM2 DRAMs. The polynomial unit is in charge of performing polynomial additions/subtractions for each TFHE logic operation, initializing bootstrapping operations, extracting samples, and conducting key-switching operations that consist of additions, logic comparisons, and Boolean logic operations. One TGSW cluster and an EP core can support one bootstrapping pipeline. As Figure~\ref{f:mat_overall_arch}(b) shows, a TGSW cluster 16 32-bit integer multipliers and 16 32-bit integer adders to support TGSW scale operations. Each TGSW cluster has only two register banks, since the memory accesses during a TGSW scale operation have strong spatial locality. The TGSW cluster can read a register bank while write the other bank concurrently. An EP core consists of an FFT core and four IFFT cores to accelerate the FFT and IFFT kernels during an EP operation, as shown in Figure~\ref{f:mat_overall_arch}(c). It has 8 register banks to serve the irregular memory accesses in FFT and IFFT kernels. An EP core also has four 32-bit integer multipliers and four 32-bit integer adders to manipulate TGSW ciphertexts during an EP operation. An FFT core is similar to an IFFT core, except its data flow. As Figure~\ref{f:mat_overall_arch}(d) highlights, an FFT core comprises an address generation unit, a twiddle factor buffer, two input/output FIFOs, and 128 butterfly cores, each of which consists of two 64-bit integer adders and two 64-bit binary shifters. The address generation unit guides butterfly cores to access the twiddle factor buffer.

\begin{table}[t!]
\centering
\footnotesize
\setlength{\tabcolsep}{3pt}
\caption{The power and area of MATCHA operating at $2GHz$.}
\begin{tabular}{|c||c|c|c|}\hline 
Name                          & Spec                                              & Power ($W$)            & Area ($mm^2$)         \\\hline\hline
TGSW                          & $\times 16$ multipliers \& adders,                & \multirow{2}{*}{0.98} & \multirow{2}{*}{0.368}  \\
cluster                       & and a 16KB, 2-bank reg. file                      &                        &                       \\\hline	
EP                            & 4 IFFT, 1 FFT, $\times 4$ multipliers \& adders,  & \multirow{2}{*}{2.87}	 & \multirow{2}{*}{1.89}  \\
core  												& and a 256KB, 8-bank reg. file                     &		                     &                       \\\hline
\textbf{Sub-total}            & $\times 8$ EP cores and TGSW clusters             &	\textbf{30.8}          & \textbf{18.06}       \\\hline\hline
polynomial									  & $\times 32$ adders \& cmps \& logic units,        & \multirow{2}{*}{2.33}  & \multirow{2}{*}{0.32} \\						
unit       									  & and a 8KB, 2-bank reg. file                       &                        &                       \\\hline														
crossbar                      &	1/2 $8\times 32/8$ NoCs (256b bit-sliced)         & 2.11                   & 0.44                  \\\hline	
SPM                           &	a 4MB, 32-bank SPM                                & 3.52                   & 3.25                  \\\hline	
mem ctrl                      &	memory controller and HBM2 PHY                    & 1.225                  & 14.9                  \\\hline\hline
\textbf{Total}                &		                                                & \textbf{39.98}         & \textbf{36.96}        \\\hline	
\end{tabular}
\label{t:t_power_area}
\end{table}

\textbf{Design Overhead}. We implemented MATCHA in RTL, and synthesized it in $16nm$ PTM process technology using state-of-the-art tools. We used CACTI to model all SPM components and register file banks. Due to its simple structure, the entire design of MATCHA can run at $2GHz$. Among various on-chip network architectures, e.g., meshs, rings, and crossbars, we selected two $8\times32$, and one $8\times8$ bit-sliced crossbars, i.e., SPM $\rightarrow$ cores/clusters, cores/clusters $\rightarrow$ SPM, and cores/clusters $\rightarrow$ cores/clusters. The hardware overhead and power consumption of MATCHA are shown in Table~\ref{t:t_power_area}. Totally, MATCHA occupies $36.96mm^2$ and consumes $39.98$ Watt. The HBM2 bandwidth is $640GB/s$.

\begin{figure}[t!]
\begin{minipage}{0.38\linewidth}
\includegraphics[width=1.1in]{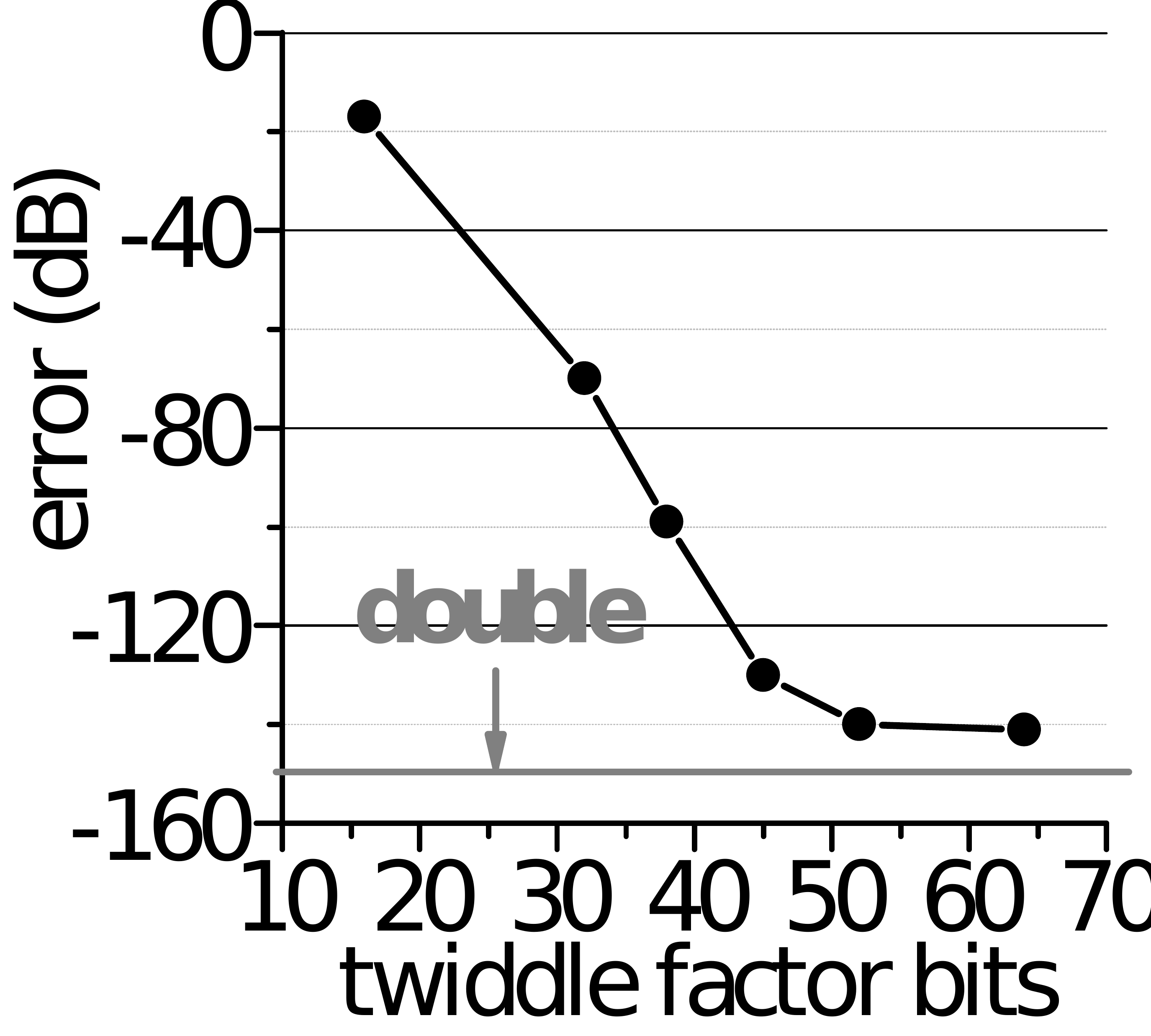}
\caption{The error of approx. FFT \& IFFT.}
\label{f:mat_noise_study}
\end{minipage}
\hfill
\begin{minipage}{0.55\linewidth}
\centering
\footnotesize
\setlength{\tabcolsep}{1pt}
\begin{tabular}{|c||c|c|}\hline
metric              & BKU~\cite{Zhou:ACCESS2018,Bourse:CRYPTO2018}        & \textbf{MATCHA} \\\hline\hline
EP                  & $\delta/2$   & $\delta/m$ \\\hline
rounding            & $\mathcal{RO}/2$   & $\mathcal{RO}/m$  \\\hline
BK                  &$3\mathcal{BK}$   & $(2^m-1)\mathcal{BK}$  \\\hline
I/FFT               & -150dB   &-141dB  \\\hline
\end{tabular}
\vspace{0.05in}
\captionof{table}{The noise comparison ($\delta$: the noise of EPs; $\mathcal{RO}$: the noise of roundings; $\mathcal{BK}$: the noise of bootstrapping keys).}
\label{t:our_tech_comparison}
\end{minipage}
\vspace{-0.05in}
\end{figure}

\textbf{Error and Noise}. The error of the polynomial multiplication result caused by approximate multiplication-less integer FFT and IFFT kernels is shown in Figure~\ref{f:mat_noise_study}. All polynomial coefficients are 32-bit integers, while we quantize the twiddle factors of FFT and IFFT with various bitwidths. With an increasing bitwidth of twiddle factors, the error caused by approximate FFT and IFFT decreases, and is similar to that generated by original double-precision floating point FFT and IFFT. With 64-bit dyadic-value-quantized twiddle factors (DVQTFs), the error caused by approximate FFT and IFFT is $\sim141dB$, which is still larger than that produced by 64-bit double-precision floating point FFT and IFFT, since the approximate FFT and IFFT perform only additions and binary shifts. At the TFHE gate level, the noise comparison between BKU and MATCHA is exhibited in Table~\ref{t:our_tech_comparison}, where BKU unrolls the bootstrapping key for two times while MATCHA unrolls that for $m$ times ($m\geq2$). With an enlarging $m$, the noise from EP and rounding operations decreases linearly, but the noise caused by bootstrapping keys increases exponentially. As a result, TFHE with a smaller $m$ can tolerate more errors caused by approximate FFT and IFFT. Based on our experiments, 38-bit DVQTFs produce no decryption failure in the test of $10^8$ TFHE gates. However, for a large $m$, e.g., $m=5$, we have to use 64-bit DVQTFs to guarantee there is no decryption failure in the same test, since the noise caused by more bootstrapping keys dominates the total noise in ciphertexts. Therefore, MATCHA adopts 64-bit DVQTFs for all approximate multiplication-less integer FFT and IFFT kernels.

\vspace{-0.05in}
\section{Experimental Methodology}
\label{s:method}

\textbf{Simulation and Compilation}: To simulate the performance of MATCHA at cycle level, we used a CGRA modeling framework, OpenCGRA~\cite{Cheng:ICCD2020}, which has been validated against multiple ASIC accelerators. OpenCGRA first compiles a TFHE logic operation into a data flow graph (DFG) of the operations supported by MATCHA, solves its dependencies, and removes structural hazards. The architecture of MATCHA is abstracted to an architecture description (AD) in OpenCGRA, which computes the latency and the energy consumption of each TFHE logic operation by scheduling and mapping the DFG onto the AD.


\begin{figure*}[t!]
\centering
\begin{minipage}{0.33\linewidth}
\centering
\includegraphics[width=2.2in]{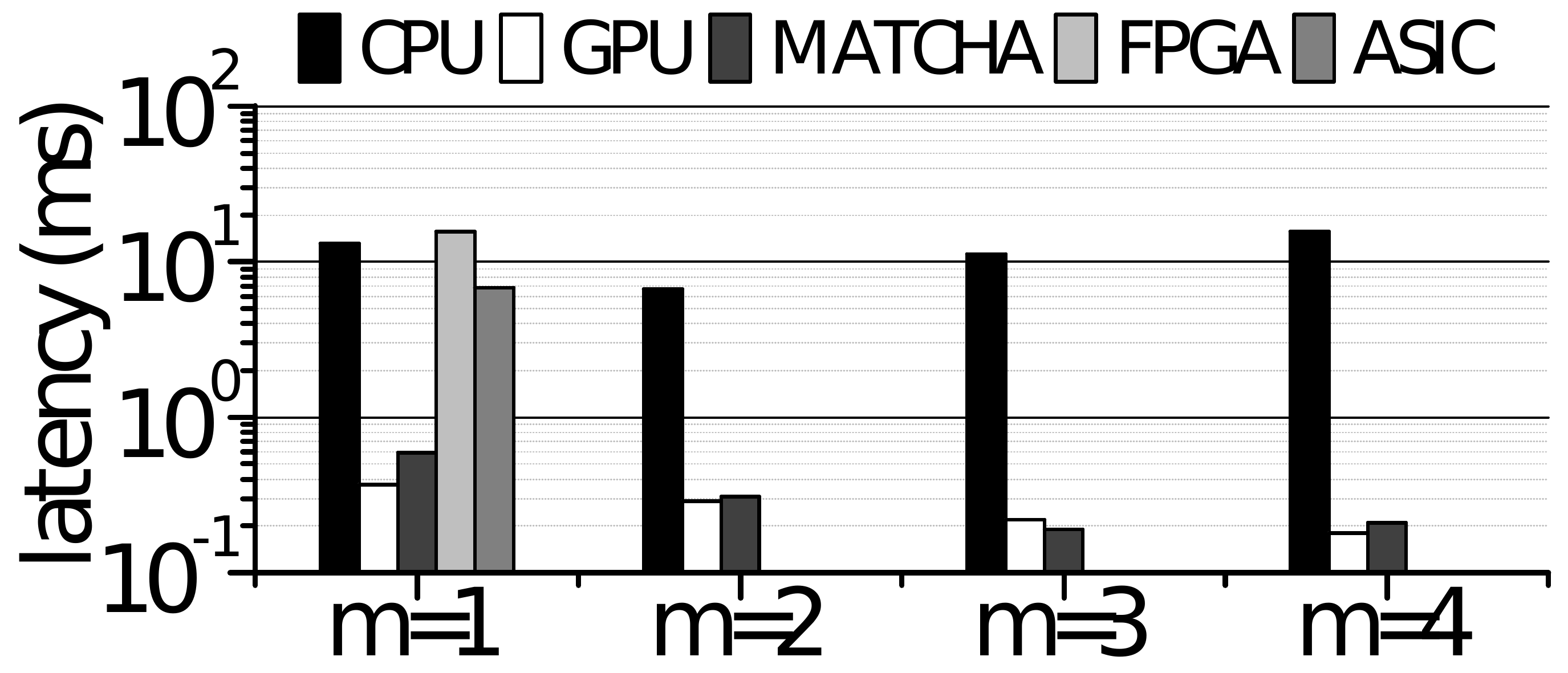}
\caption{Latency comparison.}
\label{f:mat_result_latency}
\end{minipage}
\hfill
\begin{minipage}{0.33\linewidth}
\centering
\includegraphics[width=2.2in]{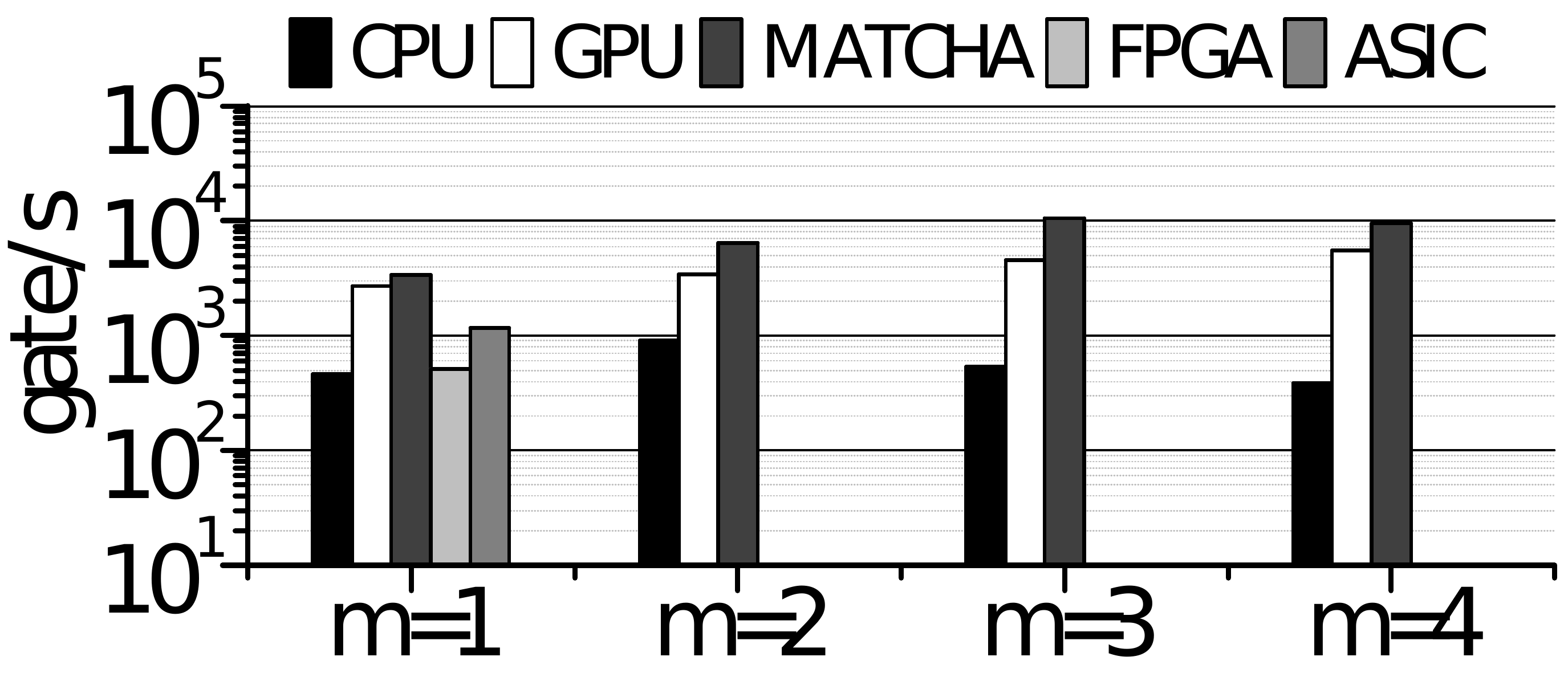}
\caption{Throughput comparison.}
\label{f:mat_result_through}
\end{minipage}
\hfill
\begin{minipage}{0.33\linewidth}
\centering
\includegraphics[width=2.2in]{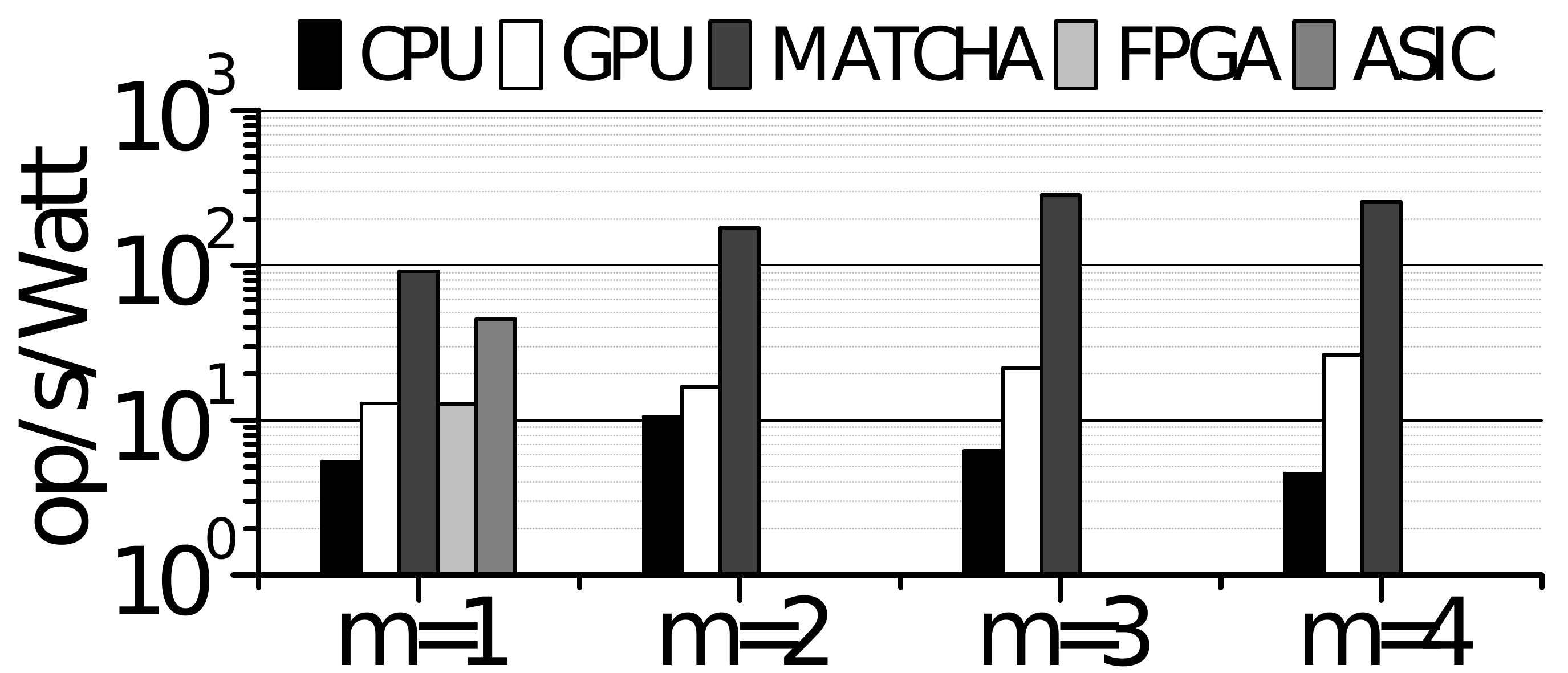}
\caption{Thrghpt/Watt comparison.}
\label{f:mat_result_power}
\end{minipage}
\vspace{-0.2in}
\end{figure*}

\textbf{Our Baselines}. We compared MATCHA against state-of-the-art CPU-, GPU-, FPGA-, and ASIC-based TFHE hardware platforms. Our CPU baseline is a 8-core $3.7GHz$ Xeon E-2288G processor executing the TFHE library~\cite{Chillotti:JC2018}, while our GPU baseline is a 5120-core Tesla-V100 GPU equipped with a 16GB HBM2 DRAM running the cuFHE library~\cite{Dai:CUFHE2018}. TFHE Vector Engine (TVE)~\cite{Serhan:SPSL2021} was implemented on a low-end ZedBoard Zynq-7000 FPGA. We implemented 8 copies of TVE on a Stratix-10 GX2800 FPGA, and used it as our FPGA baseline, since the Stratix-10 board has more resources. Because there is no existing ASIC-based design, we synthesized our FPGA baseline with the $16nm$ PTM process as our ASIC baseline. \textit{We enable BKU on CPU, GPU, and MATCHA but fix $m=1$ on FPGA and ASIC, since they do not support BKU}.

\textbf{TFHE Operations and Parameters}. We studied all TFHE logic operations including NOT, AND, OR, NAND, XOR, and XNOR, but we only report the results on NAND in Section~\ref{s:result}. This is because AND, OR, NAND, XOR, and XNOR have almost the same latency which is dominated by the bootstrapping step, while NOT has no bootstrapping at all. To maintain the standard 110-bit security, we adopt the TFHE parameters from~\cite{Chillotti:JC2018}, i.e., the polynomial degree in the ring $N=1024$, the TLWE dimension $k=1$, the basis and length for the TGSW ciphertext decomposition $Bg=1024$ and $\mathscr{l}=3$.

\vspace{-0.05in}
\section{Results and Analysis}
\label{s:result}

\textbf{Latency}. The latency comparison of a TFHE NAND gate between our various baselines and MATCHA is shown in Figure~\ref{f:mat_result_latency}. The NAND gate on CPU costs $13.1ms$, while $m=2$ reduces its latency to $6.67ms$. Aggressive BKU with an increasing $m$ cannot further reduce the NAND gate latency anymore on CPU, due to the limited number of cores, more cache conflicts, and the non-pipelined processing style. It takes only $0.37ms$ for GPU to process a NAND gate. With an enlarging $m$, GPU gradually reduces the NAND gate latency. When $m=4$, the NAND gate latency on GPU is $0.18ms$. MATCHA reduces the NAND gate latency by $13\%$ over GPU only when $m=3$, since GPU can fully use its all resources to process one TFHE gate when $m=1$ or $2$. MATCHA cannot support aggressive BKU with $m=4$ efficiently either, since it has only 8 TGSW clusters. FPGA and ASIC do not have any pipelined design or memory optimization to support BKU, and they need $>6.8ms$ to complete a NAND gate when $m=1$.

\textbf{Throughput}. The NAND gate throughput comparison between various baselines and MATCHA is shown in Figure~\ref{f:mat_result_through}. FPGA and ASIC duplicate 8 copies of the TVE~\cite{Serhan:SPSL2021}, so they support only $m=1$. By enabling aggressive BKU, even CPU ($m=2$) can achieve higher gate processing throughput than ASIC and FPGA with $m=1$. GPU and MATCHA obtain much higher throughput than ASIC, FPGA and CPU. Compared to GPU, MATCH improves the NAND gate throughput by $2.3\times$ ($m=3$), due to its pipelined architecture for aggressive BKU.

\textbf{Throughput per Watt}. The comparison of the NAND gate throughput per Watt between various baselines and MATCHA is shown in Figure~\ref{f:mat_result_power}. FPGA and ASIC consume only $\sim40W$ and $\sim26W$, and improve the NAND gate throughput per Watt by $2.4\times$ and $8.3\times$ over CPU respectively, when $m=1$. Due to the large power consumption ($>200W$) of GPU, the best throughput per Watt of GPU ($m=4$) is only about 58\% of that of ASIC. Compared to ASIC, MATCHA improves the NAND gate throughput per Watt by $6.3\times$, since it consumes only $39.98W$.

\section{Conclusion}
\label{s:con}
TFHE enables arbitrary computations with an unlimited multiplicative depth to directly occur on ciphertexts. However, TFHE gates are time-consuming and power-hungry on state-of-the-art hardware platforms. In this paper, we build MATCHA to accelerate TFHE gates. MATCHA allows aggressive bootstrapping key unrolling to process TFHE gates without decryption errors by approximate multiplication-less integer FFTs and IFFTs, and a pipelined datapath. Compared to prior CPU-, GPU-, FPGA- and ASIC-based solutions, MATCHA improves the TFHE gate processing throughput by $2.3\times$, and the throughput per Watt by $6.3\times$.

\bibliographystyle{short}
\bibliography{homo}

\begin{thebibliography}{10}
\newcommand{\enquote}[1]{``#1''}
\providecommand{\url}[1]{\texttt{#1}}
\providecommand{\urlprefix}{}

\bibitem{Becoulet:TSP2021}
A.~Becoulet and A.~Verguet, \enquote{{A Depth-First Iterative Algorithm for the
  Conjugate Pair Fast Fourier Transform},} \emph{IEEE Transactions on Signal
  Processing}, 2021.

\bibitem{Bourse:CRYPTO2018}
F.~Bourse, \emph{et~al.}, \enquote{{Fast Homomorphic Evaluation of Deep
  Discretized Neural Networks},} in \emph{Annual International Cryptology
  Conference}, 2018.

\bibitem{Brakerski:TCT2014}
Z.~Brakerski, \emph{et~al.}, \enquote{(Leveled) Fully Homomorphic Encryption
  without Bootstrapping,} \emph{ACM Transaction Computing Theory}, 6(3), July
  2014.

\bibitem{Brutzkus:ICML2019}
A.~Brutzkus, \emph{et~al.}, \enquote{{Low Latency Privacy Preserving
  Inference},} in \emph{International Conference on Machine Learning}, pages
  812--821, 2019.

\bibitem{Cheon:CEA2020}
J.~H. Cheon, \emph{et~al.}, \enquote{{Remark on the Security of CKKS Scheme in
  Practice},} Cryptology ePrint Archive, Report 2020/1581, 2020,
  \url{https://eprint.iacr.org/2020/1581}.

\bibitem{Chillotti:JC2018}
I.~Chillotti, \emph{et~al.}, \enquote{{TFHE: Fast Fully Homomorphic Encryption
  Over The Torus},} \emph{Journal of Cryptology}, 33(1):34--91, 2020.

\bibitem{Dai:CUFHE2018}
W.~Dai, \enquote{{CUDA-accelerated Fully Homomorphic Encryption Library},}
  \url{https://github.com/vernamlab/cuFHE}, 2018, worcester Polytechnic
  Institute.

\bibitem{DUCAS:ICTACT2015}
L.~Ducas and D.~Micciancio, \enquote{{FHEW: Bootstrapping Homomorphic
  Encryption in Less than A Second},} in \emph{International Conference on the
  Theory and Applications of Cryptographic Techniques}, pages 617--640,
  Springer, 2015.

\bibitem{Fan:CARCH2012}
J.~Fan and F.~Vercauteren, \enquote{{Somewhat Practical Fully Homomorphic
  Encryption},} Cryptology ePrint Archive, Report 2012/144, 2012.

\bibitem{Serhan:SPSL2021}
S.~Gener, \emph{et~al.}, \enquote{An FPGA-based Programmable Vector Engine for
  Fast Fully Homomorphic Encryption over the Torus,} \emph{SPSL: Secure and
  Private Systems for Machine Learning}, 2021.

\bibitem{Halevi:ICTACT2015}
S.~Halevi and V.~Shoup, \enquote{{Bootstrapping for HElib},} in
  \emph{International conference on the theory and applications of
  cryptographic techniques}, 2015.

\bibitem{Hoofnagle:ICTL2019}
C.~J. Hoofnagle, \emph{et~al.}, \enquote{{The European Union General Data
  Protection Regulation: What It Is \& What It Means},} \emph{Information \&
  Communications Technology Law}, 2019.

\bibitem{Liu:TECS2017}
Z.~Liu, \emph{et~al.}, \enquote{High-Performance Ideal Lattice-Based
  Cryptography on 8-Bit AVR Microcontrollers,} \emph{ACM Transactions on
  Embedded Computing Systems}, 16(4), July 2017,
  \urlprefix\url{https://doi.org/10.1145/3092951}.

\bibitem{Matsuoka:SECURITY2021}
K.~Matsuoka, \emph{et~al.}, \enquote{Virtual Secure Platform: A Five-Stage
  Pipeline Processor over {TFHE},} in \emph{{USENIX} Security Symposium}, pages
  4007--4024, 2021.

\bibitem{Ahmet:DATE2020}
A.~C. Mert, \emph{et~al.}, \enquote{{A Flexible and Scalable NTT Hardware :
  Applications from Homomorphically Encrypted Deep Learning to Post-Quantum
  Cryptography},} in \emph{Design, Automation {\&} Test in Europe Conference
  {\&} Exhibition}, 2020.

\bibitem{Toufique:HOST2020}
T.~Morshed, \emph{et~al.}, \enquote{{CPU and GPU Accelerated Fully Homomorphic
  Encryption},} in \emph{IEEE International Symposium on Hardware Oriented
  Security and Trust}, pages 142--153, 2020.

\bibitem{Oraintara:TSP2002}
S.~Oraintara, \emph{et~al.}, \enquote{Integer fast Fourier transform,}
  \emph{IEEE Transactions on Signal Processing}, 50(3):607--618, 2002.

\bibitem{Riazi:ASPLOS2020}
M.~S. Riazi, \emph{et~al.}, \enquote{{HEAX: An Architecture for Computing on
  Encrypted Data},} in \emph{ACM International Conference on Architectural
  Support for Programming Languages and Operating Systems}, 2020.

\bibitem{Feldmann:MICRO2021}
N.~Samardzic, \emph{et~al.}, \enquote{{F1: A Fast and Programmable Accelerator
  for Fully Homomorphic Encryption},} in \emph{IEEE/ACM International Symposium
  on Microarchitecture}, 2021.

\bibitem{Roy:HPCA2019}
S.~{Sinha Roy}, \emph{et~al.}, \enquote{{FPGA-Based High-Performance Parallel
  Architecture for Homomorphic Computing on Encrypted Data},} in \emph{IEEE
  International Symposium on High Performance Computer Architecture}, pages
  387--398, 2019.

\bibitem{Cheng:ICCD2020}
C.~Tan, \emph{et~al.}, \enquote{OpenCGRA: An Open-Source Unified Framework for
  Modeling, Testing, and Evaluating CGRAs,} in \emph{2020 IEEE 38th
  International Conference on Computer Design}, pages 381--388, 2020.

\bibitem{Zhou:ACCESS2018}
T.~Zhou, \emph{et~al.}, \enquote{Faster Bootstrapping With Multiple Addends,}
  \emph{IEEE Access}, 6:49868--49876, 2018.

\end{thebibliography}

\end{document}